\documentclass[11pt]{article}
\usepackage{amssymb,amsmath,amsfonts}
\usepackage{graphicx}
\usepackage{graphics}
\usepackage{eepic,epsfig}


\begin{document}

\title{Relativistic Neutron Stars: Rheological Type Extensions of the Equations of State}
\author{Alexander Balakin$^{1}$, Alexei Ilin$^{1}$, Anna Kotanjyan$^{2}$, Levon Grigoryan$^{3}$ \vspace{1cm}\\ 
\textit{$^{1}$Department of General Relativity and Gravitation, Institute of Physics,}\\
\textit{Kazan Federal University, Kremlevskaya str. 18, Kazan 420008,
Russia} \vspace{0.5cm}\\ 
\textit{$^{2}$Department of Physics, Yerevan State University,}\\
\textit{1 Alex Manoogian Street, 0025 Yerevan, Armenia} \vspace{0.5cm}\\ 
\textit{$^{3}$Institute of Applied Problems in Physics NAS RA, }\\
\textit{25 Nersessian Street, Yerevan 0014, Armenia}}
\maketitle

\begin{abstract}
Based on the Rheological Paradigm, we extend the equations of state for relativistic spherically symmetric static neutron stars, taking into consideration the derivative of the matter pressure  along the so-called director four-vector. The modified equations of state are applied to the model of a zero-temperature neutron condensate. This model includes one new parameter with the dimensionality of length, which describes the rheological type screening inside the neutron star. As an illustration of the new approach, we consider the rheological  type generalization of the non-relativistic Lane--Emden theory and find numerically  the profiles of the pressure for a number of values of the new guiding parameter. We have found that the rheological type self-interaction  makes the neutron star more compact, since the radius of the star, related to the first null of the pressure profile, decreases when the modulus of the rheological type guiding parameter grows.
\end{abstract}
\section{Introduction}

Neutron stars play the key role in Modern Relativistic Astrophysics. These compact objects are, usually, rapidly rotating and have a strong magnetic field. Relativistic pulsars are the central elements of Pulsar Physics, and form the global system of lighthouses, giving us an effective tool for investigations of the Universe structure (see, e.g., \cite{P1}-\cite{P4}). The theory of the neutron star structure is based on the analysis of the so-called {\it equations of state} (EoS); various models of the EoS are well documented and applied to the theory of pulsar structure.

The most cited is the simplest model, which follows from the Lane and Emden model (see, e.g., \cite{Weinberg} for details); it is appropriate for the non-relativistic non-rotating spherically symmetric objects with polytropic EoS.
Nowadays, the Lane--Emden model plays the role of a censor for various modified theories of star's structure: analogously with the requirement that every modified theory of gravity should have the Newtonian limit, one assumes that the theories of the star structure have to possess the appropriate Lane--Emden limit. In this sense, elaborating on a new rheological-type model of EoS in this paper, we compete below the analytical consideration by the analysis of an extension of this classical Lane--Emden model.

Relativistic models of neutron star structure use much more sophisticated versions of EoS (see, e.g., \cite{EOS1}-\cite{EOS16} and references therein). From this diversity of models, one can formally distinguish, first, the models with hot and cold (degenerated) dense matter, second, the models with pure neutron and multi-species matter (quark, meson, hadronic, etc.), third, the models with nuclear interactions between hadronic species and self-interactions inside the neutron systems. In our work, we consider one-component dense degenerated neutron matter with internal self-interaction of the rheologic type.

Another trend in the modeling of the neutron star interior is connected with the so-called modified theories of gravity (see, e.g., \cite{M1}-\cite{M9} for modifications in the framework of the Einstein--aether theory, $f(R)$ gravity, non-minimal theory). Below, we consider the rheological type EoS modifications in the standard Einstein theory of gravity.

One of the goals of the EoS modeling is connected with the problem of the neutron star compactness; to be more precise, the interest of astrophysicists is attracted to the mass/radius ratio, luminosity/radius ratio, etc. This interest was revived recently by the outstanding event encoded as GW170817 and GRB 170817A \cite{170817}: based on the observation of the binary neutron star merger, the~status of constraints on the neutron star theory was renewed. The most sound consequence of this event is that the ratio of the velocities of the gravitational and electromagnetic waves is estimated to differ from one by the quantity about $10^{-15}$. In addition, the data processing has shown that, in order to recognize the start of the final stage of the merging of two neutron stars, i.e., to recognize the last moment when the stars can be identified as individual objects,  one has to study in more detail the problem of the mass/radius ratio.
In other words, the problem of the neutron star compactness remains important and significant for both observers and theoreticians. This is one of arguments, why below we consider the neutron star model with a new EoS of rheological type, and study the mass/radius ratio as the function of a new (rheological) parameter of state.

What is the novelty of our approach to the neutron star EoS modeling? We follow the Rheological Paradigm, according to which the response of the physical system on the external actions or on the internal evolutionary processes is nonlocal in space and/or in time. In particular, when the system is spatially homogeneous, the Rheological Paradigm recommends searching for the time delay of the response and memory effects; when the system is static, the response can appear far from the domain, in which the source of perturbation is located. In the context of modeling of the EoS for static spherically symmetric neutron stars, we assume, in fact, that the profile of the pressure $P(r)$, designed with respect to the radial variable $r$, is predetermined not only by the profiles of the particle number $n(r)$ and temperature $T(r)$, but also by their directional derivatives.

What is the physical motivation of such a theory extension? Physics of complex media present us vivid examples of the so-called memory effects: one can distinguish, formally speaking,  the fading memory and the shape memory of materials  (see, e.g., \cite{Day,Shape}, respectively). One deals with the fading memory, when one considers the evolutionary processes in media, and these effects demonstrate the presence of phenomena non-local in time. The shape memory effects are associated with spatially non-local phenomena. Clearly, one can see a symmetry in the temporal and spatial aspects of non-locality. Following the idea of this symmetry, we consider below the static generalization of rheologic type phenomena based on the similarity with the known dynamic rheologic phenomena. 

Why we do believe that the idea of rheological type extension of the EoS can attract the attention of astrophysicists? There are at least three motives to think so. The first motive is that the rheological type self-interaction can make the profiles of the mass density and pressure steeper, i.e., for the same mass density at the center, for the same total mass and adiabatic index, the star radius can be smaller than the standard one. The second motive is connected with the idea to try to find new extrema in the profiles of the mass density and pressure; when the profiles are non-monotonic, one~can speak about stratification and clusterization of the neutron star interior. The third motive is that the corresponding model becomes multi-parametric, the new parameters with the dimensionality of length acquire a natural explanation, thus simplifying the fitting of guiding parameters.

The paper is organized as follows. In Section \ref{sec2}, we recall the necessary elements of classical rheological approach and the idea of its generalization. In Section \ref{sec3}, we consider the formalism of extension of the rheological approach for the case of static spherically symmetric mass configuration. We apply the established formalism to the model of cold neutron condensate in Section \ref{sec4}. As an illustration, in Subsection \ref{sec42}, we consider the rheological type generalization of the known classical Lane--Emden model. Section \ref{sec5} contains the discussion and the final section provides the conclusions.

\section{Prologue: On the Rheological Type Extension of Time\\ Dependent Constitutive Equations}\label{sec2}

In order to explain properly our approach to the theory of extended equations of state in the context of static neutron stars, we start with the analysis of the corresponding analogs given by the theory of spatially homogeneous systems admitting time delay of the response.

\subsection{Classical Constitutive Equations in Rheology}

The term {\it Rheology} was introduced by Bingham and Reiner in the 1920s; \mbox{this branch} of science studies various
aspects of behavior of media possessing elasticity, viscosity and plasticity (see, e.g., \cite{Reiner,Rabotnov}). Non-relativistic visco-elasticity started with the extension
of the standard Hooke's~law
\begin{equation}
\sigma^{\alpha \beta}(t) =  C^{\alpha \beta \mu \nu} \varepsilon_{\mu \nu}(t). \,
 \label{R0}
\end{equation}

In the standard Hooke's law  $\sigma^{\alpha \beta}$ is the stress tensor; $\varepsilon_{\mu \nu}$ describes the deformation tensor; the tensor $ C^{\alpha \beta \mu \nu}$ includes 21 elastic moduli; Greek indices correspond to spatial values (say, $1,2,3$ in the Cartesian coordinates).
The left-hand and right-hand sides are calculated at the same time $t$, thus the response of the medium was assumed to be instantaneous.
When the medium is characterized by the delay of the response, one has to write this relationship for different moments of time, for instance, as:
\begin{equation}
\sigma^{\alpha \beta}(t+\tau) =  C^{\alpha \beta \mu \nu} \varepsilon_{\mu \nu}(t), \,
\label{R13}
\end{equation}
where the parameter $\tau$ describes the time delay of the medium on the external action. If the processes develop rather slowly, one can present approximately the terms
$\sigma^{\alpha \beta}(t{+}\tau)$ as $\sigma^{\alpha \beta} {+} \tau {\dot{\sigma}}^{\alpha \beta}$; for rapid processes, the decomposition is necessary, which include derivatives of the second, third, etc. orders. In the simplest case, we recover the extended constitutive equation
\begin{equation}
\sigma^{\alpha \beta} + \tau {\dot{\sigma}}^{\alpha \beta}  =  C^{\alpha \beta \mu \nu} \varepsilon_{\mu \nu} \,
\label{R1}
\end{equation}
written by Maxwell \cite{Maxwell} (here, the dot denotes the partial derivative with respect to time). The classical theory of thermo-visco-elasticity is well-elaborated and well-documented (see, e.g., \cite{Rabotnov,JCL,Maugin99}); we~keep in mind the achievements in this sphere.

In the framework of relativistic physics Israel and Stewart have elaborated the model, indicated as Causal (Transient) Thermodynamics \cite{IsraelStewart}, which is based on the covariant extension of the Maxwell model of visco-elasticity, and uses the operator of convective derivative $D \equiv U^k \nabla_k$ instead of time derivative $\partial/\partial t$. Here, Latin indices take values $0,1,2,3$, and $U^i$ is the four-vector of macroscopic velocity of the medium. When the only scalar non-equilibrium pressure $\Pi$ appears in the model (e.g., in the spatially isotropic medium), the constitutive equation for this quantity was presented in the following form:
\begin{equation}
\Pi + \tau_{(0)} D \Pi  =  3 \zeta \Theta  + \frac{\tau_{(0)}}{2}
\Pi \left[ \Theta + D \left(\log{\frac{\tau_{(0)}}{\zeta T}}
\right) \right] \,. \label{dPI}
\end{equation}

Here, $\Theta \equiv \nabla_k U^k$ is the expansion scalar; $\zeta$ is the bulk viscosity coefficient;  $\tau_{(0)}$ is the relaxation parameter; and $T$ is the medium temperature.

In the framework of relativistic cosmology, the rheological type extensions of the equations of state for the cosmic dark energy were used in the papers  \cite{NO,BB1,BB2,BB3};  the corresponding link between the energy density $\rho$ and pressure $P$ of the dark energy was chosen as follows:
\begin{equation}
\tau D P  + P = \chi (\rho - \rho_0).\
\label{R3}
\end{equation}
The quantities $\chi$ and $\rho_0$ are the parameters of this linear model with the barotropic EoS.

\subsection{Generalization of the Time Dependent EoS with\\ Parametric Representation}

The standard barotropic equations of state form a subclass of more general two-parameter class of EoS, which is characterized by the following relationships:
\begin{equation}
P = {\cal P}(n,T) \,, \quad W = {\cal W}(n,T),\,
\label{R31}
\end{equation}
where ${\cal P}(n,T)$ and ${\cal W}(n,T)$ are given functions of two arguments, the particle number density $n$ and temperature $T$. Using the non-relativistic concept of time delay, we assume that
\begin{equation}
P(t) = {\cal P}[n(t-\tau_1),T(t-\tau_2)] \,, \quad W(t) = {\cal W}[n(t-\tau_3),T(t-\tau_4)]\,.
\label{R33}
\end{equation}

In other words, the pressure at the time moment $t$ is predetermined by the number density at the moment $t-\tau_1$ and by the temperature at the moment $t-\tau_2$ (generally, $\tau_1 \neq \tau_2$, since delay in particle migration and temperature transfer can be described by different laws). Similarly, the energy density at the time moment $t$ is predetermined by the number density at the moment $t-\tau_3$ and by the temperature at the moment $t-\tau_4$ (generally, $\tau_1 \neq \tau_3$ and $\tau_2 \neq \tau_4$). For slow varying processes, the laws (\ref{R33}) give
\begin{equation}
P(t) = {\cal P}[n(t),T(t)] - \tau_1 \frac{\partial {\cal P}}{\partial n} \dot{n} - \tau_2 \frac{\partial {\cal P}}{\partial T} \dot{T},\,
\label{R35}
\end{equation}
\begin{equation}
W(t) = {\cal W}[n(t),T(t)] - \tau_3 \frac{\partial {\cal W}}{\partial n} \dot{n} - \tau_4 \frac{\partial {\cal W}}{\partial T} \dot{T}.  \,
\label{R36}
\end{equation}

When we consider the general relativistic models, the appropriate generalizations of (\ref{R35}) and (\ref{R36}) can be obtained with replacements of partial derivatives with convective ones, $\dot{n} \to  Dn$ and $\dot{T} \to DT$. 
Below,~we use these analogs for the mathematical reconstruction of the static equations of state.

\section{The Formalism of the EoS Extension in the Framework of Static Models}\label{sec3}

\subsection{The Standard Elements of the Model: Metric, Einstein Equations, and the Equation of Hydrostatic~Equilibrium}

In this subsection, we recover the well-known elements of the theory.
We consider static spherically symmetric matter configurations, which are described by the metric
\begin{equation}\label{metrica}
ds^2=\sigma^2(r)  N(r) dt^2-\frac{dr^2}{N(r)}-r^2 \left( d\theta^2 +
\sin^2\theta d\varphi^2 \right). \,
\end{equation}

For the metric (\ref{metrica}), only four Einstein's equations are known to be
non-trivial:
\begin{equation}
\kappa W \equiv \kappa T_0^{\,0}=\frac{1-N}{r^2}-\frac{N^{\prime}}{r}, \,
\label{G1}
\end{equation}
\begin{equation}
-\kappa P_{||} \equiv \kappa T_r^{\,r}=\frac{1-N}{r^2}-\frac{N^\prime}{r}-\frac{2N\sigma^{\prime}}{r\sigma},\,
\label{G2}
\end{equation}
\begin{equation}
-\kappa P_{\bot} \equiv \kappa T_\theta^{\,\theta}= \kappa T_\varphi^{\,\varphi}=-\frac{1}{2r\sigma}\left(2\sigma
N^{\prime}+2N\sigma^{\prime}+3r\sigma^{\prime} N^{\prime}+2rN\sigma^{\prime \prime}+r\sigma N^{\prime \prime}\right).
\label{G3}
\end{equation}

The prime denotes the derivative with respect to the radial variable $r$.
Here, we introduced the energy density scalar $W$ of the matter, longitudinal pressure $P_{||}$ and transversal pressure
$P_{\bot}$.

The direct consequence of (\ref{G1}) is the expression of $N(r)$ through the integral of the function $W(r)$ (when its profile is obtained):
\begin{equation}
N(r) = 1- \frac{2G M(r)}{r} \,, \quad  M(r) = \frac{4\pi}{c^2} \int_0^r {\tilde{r}}^2 d\tilde{r} W(\tilde{r}), \,
\label{G4}
\end{equation}
where $M(r)$
is usually associated with the mass inside the sphere of the radius $r$.
The function $\sigma(r)$ can be formally found by integration of the equation
\begin{equation}
\frac{\sigma'}{\sigma} = \frac{r \kappa}{2N} \left(W+P_{||} \right) \,
\label{G6}
\end{equation}
(when the profiles $W(r)$, $P_{||}(r)$ and $N(r)$ are known).
As usual, instead of (\ref{G3}), we consider the consequence of the Bianchi identity
$\nabla_k T^k_{i} = 0$, which in our context reads
\begin{equation}
P^{\prime}_{||} + \frac{2}{r} \left(P_{||}-P_{\bot} \right) + (W+P_{||})\left(\frac{N^{\prime}}{2N} + \frac{\sigma^{\prime}}{\sigma} \right) =0. \,
\label{G8}
\end{equation}

Excluding the metric coefficients, we obtain the well-known standard equation of hydrostatic equilibrium:
\begin{equation}
P^{\prime}_{||} + \frac{2}{r} \left(P_{||}-P_{\bot} \right) + \frac{G}{r^2 c^4} \left(Mc^2+4\pi r^3 P_{||} \right) \left(1-\frac{2GM}{r c^2} \right)^{-1} \left(W+P_{||} \right) =0 \,.
\label{G9}
\end{equation}

In other words, (\ref{G9}) is the key equation of the problem.  It will be solvable, when we add two equations of state, which connect $P_{||}$ and $P_{\bot}$ with $W$, directly in the scheme of barotropic equations of state, or in the parametric form as functions of the temperature $T(r)$ and particle density $n(r)$.

\subsection{Extended Equations of State for the Static Configurations}

\subsubsection{Convective and Directional Derivatives}

The unit time-like velocity four-vector $U^i$, which characterizes the medium motion, can be used as a tangent four-vector  for reconstruction of the global reference frame with world-lines satisfying the equation $dx^i/ds = U^i$. These world-lines form the system of time-like geodesics, when $U^i$ is normalized by unity $U^kU_k = 1$ and satisfies the requirements
\begin{equation}
  \frac{d^2 x^i}{ds^2} + \Gamma^i_{jk} \frac{d x^j}{ds}\frac{d x^k}{ds} =0  \ \ \ \Rightarrow \ \ \ U^k \nabla_k U^i = 0.  \,
\label{D1}
\end{equation}

The corresponding differential operator $D \equiv U^k \nabla_k$, the convective derivative, is widely used in relativistic theory of media, and the equation $DU^i=0$ can be interpreted as the absence of acceleration in the medium velocity flow.

Analogously with the convective derivative, one can introduce the so-called directional derivative ${\cal D} \equiv {\cal R}^k \nabla_k$, where the unit space-like four-vector ${\cal R}^k$ indicated as a director is defined as follows:
\begin{equation}
{\cal R}^k \nabla_k {\cal R}^i =0 \,, \quad {\cal R}^k{\cal R}_k = -1 \,.
\label{D2}
\end{equation}

When we deal with the metric (\ref{metrica}), the appropriate solutions for $U^i$ and ${\cal R}^i$ are, respectively,
\begin{equation}
U^i = \delta^i_0 \frac{1}{\sigma \sqrt{N}} \,, \quad {\cal R}^i  = \delta^i_r \sqrt{N}. \,
\label{D3}
\end{equation}

The scheme of generalization is very simple: we replace the convective derivatives with the directional~ones.

\subsubsection{Static Analogs of the Rheologically Extended Constitutive Equations}

Now, we introduce the  static spherically symmetric analogs of the rheological type equation of state. Our general ansatz is the following:
\begin{equation}
P_{||} = {\cal P}_{||}[n,T)] - \gamma_1 \frac{\partial {\cal P}_{||}}{\partial n} {\cal D}n - \gamma_2 \frac{\partial {\cal P}_{||}}{\partial T} {\cal D}T,\,
\label{R351}
\end{equation}
\begin{equation}
P_{\bot} = {\cal P}_{\bot}[n,T] - \tilde{\gamma}_1 \frac{\partial {\cal P}_{\bot}}{\partial n} {\cal D}n - \tilde{\gamma}_2 \frac{\partial {\cal P}_{\bot}}{\partial T} {\cal D}T,\,
\label{R37}
\end{equation}
\begin{equation}
W = {\cal W}[n,T] - \gamma_3 \frac{\partial {\cal W}}{\partial n} {\cal D}n  - \gamma_4 \frac{\partial {\cal W}}{\partial T} {\cal D}T.  \,
\label{R361}
\end{equation}

Here, ${\cal D} \equiv {\cal R}^k \nabla_k$; the scalars $\gamma_1$, $\gamma_2$, $\gamma_3$, $\gamma_4$, $\tilde{\gamma}_1$, $\tilde{\gamma}_2$  play the roles of specific correlation radii; they are the analogs of the relaxation parameters in the time dependent models. Clearly, when the pressure is isotropic, i.e., $P_{||}=P_{\bot}=P$, we obtain only two extended equations.

\section{The Model of Cold Isotropic Neutron Condensate}\label{sec4}

\subsection{Extended Equation of Hydrostatic Equilibrium: The General Relativistic Model}

Let us consider the model, in which the temperature is equal to zero, $T(r)=0$, the directional derivative of the temperature is vanishing, ${\cal D}T=0$, and  the pressure is of Pascal type, $P_{||}=P_{\bot}=P(r)$.
For the standard model, the equation of state can be presented in the well-known one-parameter form (see, e.g., \cite{Weinberg,EOS1,OV}):
\begin{equation}
P \to {\cal P} = \frac{1}{32} n^{*}mc^2 \left[\sinh{\xi} - 8 \sinh{\frac{\xi}{2}} + 3\xi \right],\,
\label{Q1}
\end{equation}
\begin{equation}
W \to {\cal W} = \frac{3}{32} n^{*}mc^2 \left[\sinh{\xi} - \xi \right],\,
\label{Q2}
\end{equation}
\begin{equation}
n \to n^{*}\sinh^3{\frac{\xi}{4}} \,, \quad n^{*} = \frac{8\pi m^3 c^3}{3h^3}. \,
\label{Q3}
\end{equation}

The state parameter $\xi$ is assumed to be the function of the radial variable $r$.
The critical value $n^{*}$ of the particle number density $n(r)$ corresponds to the case, when the Fermi-momentum $p_{F}$ is equal to the quantity $mc$:
\begin{equation}
\frac{p_{F}}{mc} \equiv \frac{h}{mc} \left(\frac{3n}{8\pi} \right)^{\frac13} = \left(\frac{n}{n^{*}} \right)^{\frac13}. \,
\label{N1}
\end{equation}

This parametrical representation of the EoS is known to be valid, when the energy density at the center, $W(0)$ is less than a some critical value $W_{\rm crit}$; when $W(0)>W_{\rm crit}$, the neutron gas becomes instable and can be converted into the multi-component hadronic plasma (see, e.g., \cite{P4,EOS1,EOS2} for details).

Let us consider the
{\it extended} model, based on the following assumption: $\gamma_3=0$, so that the energy density scalar is described by the same Equation (\ref{Q2}). Then, keeping in mind the formula for directional derivative ${\cal D} = {\cal R}^i \nabla_i$ with  (\ref{D3}), we can present the equation for the pressure in the following form
\begin{equation}
P(r) +  \gamma_1(r) \frac{\partial {\cal P}}{\partial n}\sqrt{N(r)} \ n^{\prime}(r)   = {\cal P}(r),   \,
\label{Q4}
\end{equation}
where the function ${\cal P}$ is given by the Formula (\ref{Q1}).
In terms of $\xi(r),$ the last equation can be written as~follows:
\begin{equation}
P  = \frac{1}{32} n^{*}mc^2 \left[\sinh{\xi} - 8 \sinh{\frac{\xi}{2}} + 3\xi  - \Gamma \ \frac{d\xi}{dr} \left(\cosh{\xi} - 4 \cosh{\frac{\xi}{2}} + 3 \right)\right].   \,
\label{Q5}
\end{equation}

Here, we introduce the function $\Gamma(r) = \gamma_1(r) \sqrt{N(r)}$ instead of $\gamma_1(r)$.
The equation of Hydrostatic Equilibrium (\ref{G9}) now takes the form
\begin{equation}
P^{\prime}  + \frac{G}{r^2 c^4} \left(Mc^2+4\pi r^3 P \right) \left(1-\frac{2GM}{r c^2} \right)^{-1} \left(W+P\right) =0. \,
\label{G90}
\end{equation}

Thus, in order to find the profile $\xi(r),$ we have to solve (\ref{G90}) into which we insert the function $P(r)$ extracted from (\ref{Q5}), the function $W(r)$ given by (\ref{Q2}) and the mass $M(r)$ given by
\begin{equation}
M(r)=\frac38 \pi n^*m \int_0^r\tilde{r}^2 d\tilde{r} (\sinh{\xi}-\xi)(\tilde{r}), \,
\label{G909}
\end{equation}
obtained from (\ref{G4}). Then, we can extract $M(r)$ from (\ref{G90})
\begin{equation}
\frac{GM(r)}{c^2}=\frac{r^2\left[P^{\prime}+ \frac{4\pi G}{c^4} r P ({\cal W}+P) \right]}{\left[2rP^{\prime} - ({\cal W}+P) \right]} \,
\label{G55}
\end{equation}
and exclude the integral by differentiation of (\ref{G55}). As a result, we obtain the equation
\begin{equation}
\frac{3\pi G n^*m }{8c^2} (\sinh{\xi}-\xi)=  \frac{1}{r^2}\frac{d}{dr}\left\{\frac{r^2\left[\frac{dP}{dr}+ \frac{4\pi G}{c^4} r P ({\cal W}+P) \right]}{\left[2r \frac{dP}{dr} - ({\cal W}+P) \right]} \right\} \,
\label{G551}
\end{equation}
into which we have to insert $P(\xi(r))$ given by (\ref{Q5}). Clearly, it is the very sophisticated nonlinear differential equation of the third order for the quantity $\xi(r)$. This equation admits the numerical study only, and we optimistically hope to provide this analysis in a separate work. However, in order to demonstrate the capacities of the rheological type model under consideration, we complete our theoretical modeling by the study of the generalization of the classical Lane--Emden equation.

\subsection{Rheological Type Generalization of the Non-Relativistic Lane--Emden Equation}\label{sec42}

In the context of study of the Fermi gas with $T=0$, the term non-relativistic means that the Fermi-momentum $p_{F}$ is much less than the quantity $mc$, and, respectively, the particle number density is much less than the critical value:
\begin{equation}
\frac{p_{F}}{mc} = \left(\frac{n}{n^{*}} \right)^{\frac13} << 1. \,
\label{N1}
\end{equation}

In accordance with (\ref{Q2}), (\ref{Q3}), (\ref{Q5}), we obtain now that $\xi \to 0$, and the following approximation is~valid:
\begin{equation}
n \simeq \frac{n^{*}}{64}\xi^3 \left(1 + \frac{\xi^2}{32} + ... \right) \,, \quad W \simeq \frac{1}{64} n^{*}mc^2 \xi^3 \left(1+ \frac{\xi^2}{20} + ...\right),\,
\label{NQ2}
\end{equation}
\begin{equation}
P  \simeq \frac{1}{2^{10}} n^{*}mc^2 \xi^5 \left(\frac15 - \Gamma \frac{\xi^{\prime}}{\xi} \right).   \,
\label{NQ5}
\end{equation}

Taking into account that $P<<W$ in the non-relativistic limit, one can rewrite the Equation (\ref{G55}) in the quasi-Newtonian form:
\begin{equation}
-\frac{GM(r)}{c^2}= \frac{r^2}{32} \left[ (\xi^2)^{\prime} - 2 \Gamma_0 (\xi \xi^{\prime \prime}+ 4 {\xi^{\prime}}^2) \right],
\label{N55}
\end{equation}
where we have changed the function $\Gamma(r)$ by the constant $\Gamma_0$. Differentiation of this equation with respect to $r$ gives the equation of the third order, which can be rewritten as the extended Lane--Emden equation with $\gamma = 5/3$ \cite{Weinberg}
\begin{equation}
\frac{1}{x^2} \frac{d}{dx}\left( x^2 \frac{d \Theta}{dx} \right) +
\Theta^{3/2}
=\frac{1}{2}\Gamma_{*}\left[\Theta^{ \prime \prime \prime}+\frac{3(2\Theta\Theta^{\prime}\Theta^{\prime \prime}-
{\Theta^{\prime}}^3)}{2\Theta^2}+\frac{1}{x}\left(2\Theta^{\prime \prime}+
3 \frac{{\Theta^{\prime}}^2}{\Theta}\right)\right]. \,
\label{LE4}
    \end{equation}

In this equation, we have introduced the generalized Lane--Emden function $\Theta(x)$:
\begin{equation}
\Theta(x) = \frac{1}{16}\xi^2 \left(\frac{n^*}{n(0)}\right)^{2/3}, \,
\label{LE2}
\end{equation}
the dimensionless radial variable $x$:
\begin{equation}
x = r \left(\frac{n(0)}{n^*}\right)^{1/6} \frac{\sqrt{8\pi G m n^{*}}}{c}, \,
\label{LE1}
\end{equation}
and the dimensionless parameter $\Gamma_*$:
\begin{equation}
\Gamma_{*} = \frac{1}{c} \Gamma_0 \sqrt{32\pi G m n^{*}} \left(\frac{n(0)}{n^*}\right)^{1/6}.  \,
\label{LE5}
\end{equation}

\subsubsection{Behavior of the Function $\Theta(x)$ near the Center}

In terms of the generalized Lane--Emden function $\Theta$ the function $n(r)$ transforms into
\begin{equation}
n(r) \rightarrow n(x) = n(0)\Theta^{3/2}(x), \,
\label{LE3}
\end{equation}
thus providing the classical condition at the center, $\Theta(0)=1$.
We can now decompose the function $\Theta(x)$ in the vicinity of the center
\begin{equation}
\Theta(x) \to 1 + \alpha x + \beta x^2 + \gamma x^3 + \sigma x^4 + \dots     \,
\label{decomp1}
\end{equation}
and, to link the parameters $\alpha$, $\beta$, $\gamma$, $\sigma$
using the requirement that $\Theta(x)$, which satisfies the equation (\ref{LE4}), is regular at $x \to 0$.
In order to untwine the obtained coupled system, we consider the $\Gamma_{*}$ to be a small parameter, and put $\sigma = 1/80$, the value, which corresponds to the Lane--Emden parameter for $\Gamma_{*}=0$. Then, we obtain the reduced series of coupled relations:
$$
2\alpha=\frac{1}{2}\Gamma_{*}\left(4\beta+3\alpha^2\right), \,
$$
$$
6\beta+1=\frac{1}{2}\Gamma_{*}\left(18\gamma + 18\alpha\beta - \frac{9}{2}\alpha^3 \right), \,
$$
\begin{equation}
4\gamma+\frac{1}{2}\alpha=\frac{1}{2}\Gamma_{*}\left(16\sigma + 2\alpha^4+6\alpha\gamma+8\beta^2-10\alpha^2\beta\right). \, \dots
\label{decomp5}
\end{equation}

Then, the leading order terms in the decomposition (\ref{decomp1}) form the function     \begin{equation}
\Theta(x) \simeq 1-\frac{1}{6}\Gamma_{*}x - \frac{1}{6}x^2 + \frac{53}{720}\Gamma_{*}x^3 + \frac{1}{80}x^4 + \dots. \,
\label{dec3}
\end{equation}

Clearly, when $\Gamma_{*}=0$, we recover the well-known Lane--Emden decomposition \cite{Weinberg}. As for the conditions at the regular center, we obtain immediately from (\ref{dec3}) that
\begin{equation}
\Theta(0)=1; \quad \Theta^{\prime}(0)=-\Gamma_{*}/6; \quad \Theta^{\prime \prime}(0)=-1/3. \,
\label{dec67}
\end{equation}

When $\Gamma_{*}=0$, the function $\Theta(x)$ has the maximum at $x=0$, since $\Theta^{\prime}(0)=0$ and $\Theta^{\prime \prime}(0)= -1/3$; when $\Gamma_{*}\neq 0$ the maximum disappears.

\subsubsection{Behavior of the Pressure $P(x)$}

In terms of the function $\Theta(x),$ the pressure $P$ can be reconstructed, using (\ref{NQ5}) and (\ref{LE2})--(\ref{LE5}), as~follows:
\begin{equation}
  P(x) = P_{*} \ \Theta^{3/2}(x)  \left[\Theta(x)- \frac54 \Gamma_{*} \Theta^{\prime}(x) \right], \,
 \label{P1}
\end{equation}
where the convenient parameter $P_{*}$ is
\begin{equation}
  P_{*} = \frac15 n^{*}mc^2 \left(\frac{n(0)}{n^*}\right)^{5/3}. \,
\label{P2}
\end{equation}

Clearly, the pressure is regular at the center, since
\begin{equation}
P(0) = P_{*}  \left(1-\frac{5}{24} \Gamma^2_{*} \right) \,, \quad P^{\prime}(0) = - \frac{5}{96}P_{*} \Gamma^3_{*}.  \,
\label{P3}
\end{equation}

As usual, the radius of the star, $R_0$ can be found as
\begin{equation}
R_0 = x_1  \left(\frac{n^*}{n(0)}\right)^{1/6} \frac{c}{\sqrt{8\pi G m n^{*}}}, \,
\label{P4}
\end{equation}
where $x_1$ is the first null of the function $P(x)$.
When $\Gamma_*=0$, we have to analyze the equation $\Theta(x){=0}$, and obtain the value $x_1 \simeq 3.65$ known from the Lane--Emden theory for the polytropic star with $\gamma=5/3$ (see, e.g., \cite{Weinberg}).
When $\Gamma_* \neq 0$, the equation $P(x)=0$ (see (\ref{P1})) splits into two equations: $\Theta(x)=0$ and $\Theta(x)= (5/4)\Gamma_{*} \Theta^{\prime}(x)$.
The function $\Theta(x)$ is positive and $\Theta^{\prime}(x)$ is negative on the interval $0<x<x_1$, thus there exists a new branch of solutions, when $\Gamma_{*}<0$, and this new branch does not appear, when $\Gamma_{*}>0$.
For the case $\Gamma_*<0$, the radius of the star decreases in comparison with the Lane--Emden value (see Figure 1).

\begin{figure}[h]
    \includegraphics[width=115mm,height=100mm]{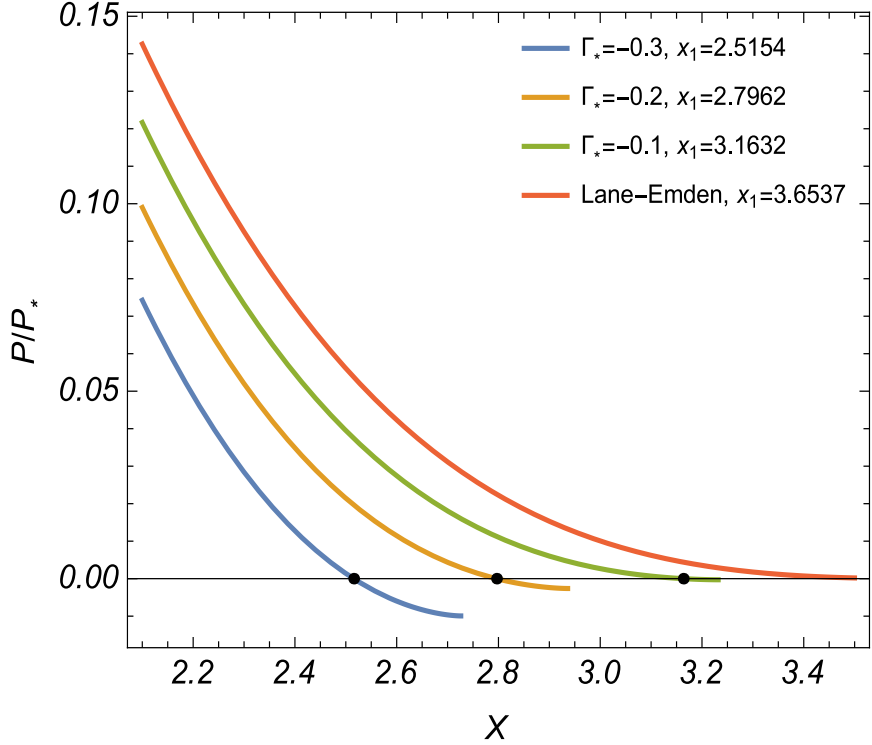}
    \caption{{\small Plot of the function $P(x,\Gamma_*)/{P_{*}}$ in the domain of the first nulls; this plot depicts the profiles of the reduced pressure as the function of the dimensionless rheological parameter $\Gamma_*$. For negative $\Gamma_*$, the radius of the star, predetermined by the condition $P(x_1){=}0$, becomes smaller than the radius predicted by the Lane--Emden theory; for positive $\Gamma_*$ there are no roots of the function $P(x,\Gamma_*)$. }}
\end{figure}

\subsubsection{The Mass/Radius Ratio}

The total mass of the star can be obtained from (\ref{N55}) at $r=R_0$. If we use the relation (\ref{LE1}) for $x=x_1$, extract the value $n(0)$ from this equation, and then put it into (\ref{N55}), we obtain the so-called mass/radius ratio:
\begin{equation}
M(R_0)R^3_0 = \frac{9\pi^2 \hbar^6}{320 m^8 G^3} {\cal F}(x_1) \,, \quad {\cal F}(x_1) = x^5_1 \left[\frac54\Gamma_* \Theta^{\prime \prime}(x_1)- \Theta^{\prime}(x_1) \right]. \,
\label{N559}
\end{equation}

The Lane--Emden ratio coincides with (\ref{N559}), when $\Gamma_*=0$.
The values ${\cal F}(x_1)$ are visualized in the Table \ref{tab1} for several values of the guiding parameter $\Gamma_*$.

\begin{table}[h]
\caption{The value $x_1$ relates to the first zero of the function $P(x)$; it determines the radius of the object and depends on the value of the guiding parameter $\Gamma_*$. The function ${\cal F}(x_1)= x_1^5\left[(5/4)\Gamma_* \Theta^{\prime \prime}(x)- \Theta^{\prime}(x)\right]$ enters the mass/radius ratio (\ref{N559}). The case $\Gamma_*=0$ corresponds to the Lane--Emden model. For the fixed mass $M(R_0)$, the radius of the object, $R_0$, decreases, if the modulus of the guiding parameter, $|\Gamma_*|$, grows.}\label{tab1}
\centering
\begin{tabular}{ccc}

   $ \ \ \Gamma_* = 0$ \ \ \ \ \ \ \ & $ \ \ x_1 = 3.6537 $ \ \ & \ \ ${\cal F}(x_1)= 132.384$ \ \  \\
    $ \ \ \Gamma_* = - 0.1$ \ \  & $ \ \ x_1 = 3.1632 $ \ \  & $ \ \  {\cal F}(x_1)= 197.746 $ \ \  \\
    $ \ \ \Gamma_* = -0.2$ \ \  &  \ \ $ x_1 = 2.7962 $ \ \  & \ \ ${\cal F}(x_1)= 125.438 $ \ \  \\
$ \ \ \Gamma_* = -0.3$ \ \ & \ \  $ x_1 = 2.5154 $ \ \ & \ \  ${\cal F}(x_1)= 84.266 $ \ \  \\
\end{tabular}
\end{table}

\section{Conclusions}\label{sec5}

The main goal of this work is to establish the formalism of rheological type extension of the theory, which describes the interior of the static spherically symmetric relativistic compact objects. The~key element of this extended model is formed by the rheologically generalized equations of state. For~general cases, when the pressure tensor has a non-Pascal structure, the corresponding equations of state are obtained to be of the form  (\ref{R351})--(\ref{R361}). The extended equation of state for the neutron condensate with the Pascal type pressure is simplified to look like (\ref{Q4}). When we deal with the extension of the non-relativistic Lane--Emden model, the modified EoS is presented by (\ref{P1}) with $\Theta(x)$ found from (\ref{LE4}). In our model, the Equation of Hydrostatic Equilibrium has formally speaking the standard form (\ref{G9}); however, it converts into the sophisticated nonlinear differential equation of the third order, when we use the rheologically extended equations of state (see (\ref{G551})).

In order to visualize analytically the capacities of the presented new model, we have considered the rheological type extension of the well-known Lane--Emden theory (see Subsection IVB). \mbox{This extended} model contains one dimensionless guiding parameter $\Gamma_*$. The profiles of the pressure $P(r,\Gamma_*)$ as the function of radial variable and of this guiding parameter demonstrate the following interesting features. First of all, the presence of the parameter $\Gamma_*$ allows the object to be  extra-compact. Indeed, when $\Gamma_*$ is negative, the first null of the pressure profile, which gives the radius of the object $R_0$, appears at the value of the radial variable less than for classical Lane--Emden prediction (see Figure 1). For~positive  $\Gamma_*$, there are no new nulls of the equation $P(r)=0$. This~fact can be interpreted as follows. Contrarily to the time delay in dynamic systems, for which the response appears later than the external impact took place, the rheological type ``response'' of the static system is localized closer to the center, in which the variation of the mass density occurred.
The~second interesting feature is that, when the total mass of the object is fixed, and the modulus of the guiding parameter, $|\Gamma_*|$, increases, the radius of the neutron star becomes smaller (see (\ref{N559}) and Table \ref{tab1}).
The~third interesting feature is that the mass density and pressure profiles become non-monotonic. Indeed, according to the decomposition (\ref{dec3}), valid near the center for small $\Gamma_*$, there is the maximum at $x_{\rm max} \simeq -\Gamma_*/2$ characterized by the value $\Theta_{\rm max} \simeq 1+ \Gamma^2_*/24$, which appears in the profile of $\Theta(x)$, if the guiding parameter
$\Gamma_*$ is negative. When $\Gamma_* =0$, the maximum is placed at the center $x=0$, so that the profile of $\Theta(x)$ remains monotonic for $x>0$. The profile of the pressure behaves similarly.
We emphasize once again that these features appear, since in the rheological type equation of state the pressure depends not only on the mass density, but also on the derivative of this quantity.

For the relativistic model with a strong gravitational field, arbitrary rheological parameter $\Gamma$, and~with the pressure expressed in terms of $\xi(r)$ using (\ref{Q5}), the Equation of Hydrostatic Equilibrium~(\ref{G551}) supplemented by (\ref{Q2}) becomes a nonlinear differential equation of the third order for the auxiliary function $\xi(r)$. In the classical theory of the neutron stars, when $\Gamma=0$, the corresponding differential equation is of the second order. Thus, in the rheologic type model, we are faced with much more sophisticated mathematical problem, which guarantees the presence of new solutions, and~probably, a number of surprises. In the near future, we hope to fulfil full-format numerical simulations for the nonlinear differential equation of the third order~(\ref{G551}). We~hope that in the full-format version of the model the extrema of the profiles $\xi(r)$, $P(r)$, $W(r)$ will be found, thus describing the clusterization of the neutron star interior and modification of the mass/radius ratio. We also intend to consider the generalization of the nuclear EoS taking into account the rheological terms in
(\ref{R351})--(\ref{R361}). We are hoping that this work will bring us many new interesting results.

\section*{Acknowledgments}

The work was partially supported by the Program of Competitive Growth of Kazan Federal University and the grant  18T-1C355 of the Committee on Science Ministry of Education and Science of the Republic of Armenia.

\end{document}